\documentclass[manuscript,astrosymb]{emulateapj}

\usepackage[normalem]{ ulem }
\usepackage{soul}
\usepackage{xcolor}

\usepackage{amsmath}
\def\Rearth{R_\oplus}
\def\Mearth{M_\oplus}

\graphicspath{{./}{figures/}}

\received{}
\revised{}
\accepted{\today}

\shorttitle{Irradiated ocean planets bridge super-Earth and sub-Neptune populations}
\shortauthors{Mousis et al.}

\begin{document}

\def\Rearth{R_\oplus}
\def\Mearth{M_\oplus}

\title{Irradiated ocean planets bridge super-Earth and sub-Neptune populations}
\email{olivier.mousis@lam.fr}

\author{Olivier Mousis, Magali Deleuil, Artyom Aguichine}
\affil{Aix Marseille Univ, CNRS, CNES, LAM, Marseille, France}
\author{Emmanuel Marcq}
\affil{LATMOS/CNRS/Sorbonne Universit\'e/UVSQ, 11 boulevard d'Alembert, Guyancourt, F-78280, France}
\author{Joseph Naar}
\affil{Aix Marseille Univ, CNRS, CNES, LAM, Marseille, France}
\affil{LATMOS/CNRS/Sorbonne Universit\'e/UVSQ, 11 boulevard d'Alembert, Guyancourt, F-78280, France}
\author{Lorena Acu\~na Aguirre}
\affil{Aix Marseille Univ, CNRS, CNES, LAM, Marseille, France}
\author{Bastien Brugger}
\affil{Department of Astronomy, Cornell University, Ithaca, NY 14853, USA}
\author{Thomas Gon\c calves}
\affil{Aix Marseille Univ, CNRS, CNES, LAM, Marseille, France}

\begin{abstract}
Small planets ($\sim$1--3.9 $\Rearth$) constitute more than half of the inventory of the 4000-plus exoplanets discovered so far. Smaller planets are sufficiently dense to be rocky, but those with radii larger than $\sim$1.6 $\Rearth$ are thought to display in many cases hydrogen/helium gaseous envelopes up to $\sim$30\% of the planetary mass. These low-mass planets are highly irradiated and the question of their origin, evolution, and possible links remains open. Here we show that close-in ocean planets affected by greenhouse effect display hydrospheres in supercritical state, which generate inflated atmospheres without invoking the presence of large hydrogen/helium gaseous envelopes. We present a new set of mass-radius relationships for ocean planets with different compositions and different equilibrium temperatures, which are found to be well adapted to low-density sub-Neptune planets. Our model suggests that super-Earths and water-rich sub-Neptunes could belong to the same family of planets, i.e. hydrogen/helium-free planets, with differences between their interiors simply resulting from the variation in the water content. 
\end{abstract}


\section{Introduction}

With radii ranging between those of the Earth (1 $\Rearth$) and Neptune ($\sim$3.9 $\Rearth$), small planets constitute more than half of the inventory of the 4000-plus exoplanets discovered so far\footnote{https://exoplanetarchive.ipac.caltech.edu}. Smaller planets are sufficiently dense to be rocky, but those with radii larger than $\sim$1.6 $\Rearth$ are thought to display large amounts of volatiles, including in many cases hydrogen/helium gaseous envelopes up to $\sim$30\% of the planetary mass \citep{Lo12,We14,Ro15}. With orbital periods less than 100 days, these low-mass planets are highly irradiated and their origin, evolution, and possible links are still debated \citep{Ve17,Do17,Ji18,Ze19}. 

The composition of the small planet population is often assessed by comparing their mass and radius measurements, when available, to theoretical mass/radius relationships. While super-Earths appear compatible with more or less differentiated rocky planets, sub-Neptunes fall near non physical curves of planets composed of pure water, suggesting instead a solid core surrounded by a hydrogen/helium gaseous envelope up to $\sim$30\% of the planetary mass \citep{Fu17,Fu18,Lo12,Ro15,Ve17,We14,Do17,Ji18,Ze19}. However, even if the existence of sub-Neptunes with low metallicities (i.e. with hydrogen-rich atmospheres) is supported by spectroscopic observations \citep{Be19a}, H$_2$O absorption bands have also been detected in the exo-Neptunes HAT-P-26b \citep{Ma19} and K2-18b \citep{Be19b}. 

There is no clear consensus on how these small exoplanets populations might have formed. In-situ formation faces number of issues  \citep{Og15} and migration appears as the most likely scenario. In that case, planetary embryos or building blocks might have formed in the cold regions of the protoplanetary disk and migrated inward. As a result of their formation location, they should contain a significant fraction of water ice. This leads to reconsider the possible existence of massive water-rich planets, as initially suggested by \cite{Le04}. Water-rich worlds (Europa, Titan, Enceladus, Pluto, etc) are ubiquitous in our solar system, and the building blocks of Neptune and Uranus are also supposed to be water rich \citep{Mo18}. However none of the existing mass/radius relationships fully explores the physical properties of such ocean planets, i.e. planets displaying a significant fraction of liquid water ($\sim$50 wt\%), especially in the domain of high surface temperature. For instance, \cite{Ze14} and \cite{Th16} pointed out the possible presence of supercritical water in ocean planets, but both were lacking in an atmospheric prescription. On the other hand, a recent investigation of mass-radius relationships for water-rich terrestrial planets (up to 5 wt\% of water) endowed with water-dominated steam atmospheres suggests that such planets present radii larger those of planets with water in liquid or solid phases  \citep{Tu19}. However, this study needs further reassessment because the water layers of the most hydrated planets considered encompass pressure-temperature ranges corresponding to those of a supercritical fluid, which is not implemented in the used atmospheric model.

Here, to investigate the role of irradiation on ocean planets, we use a combination of two one-dimensional models, i.e. a fully differentiated planet interior model \citep{Br17} and a steam atmosphere model \citep{Ma12,Ma17} connected at a 1000-bar pressure, both using a supercritical water equations of state (EOS). We show that close-in ocean planets \citep{Le04} affected by greenhouse effect display hydrospheres in supercritical state, which generate inflated atmospheres without invoking the presence of large H/He gaseous envelopes. We derive a new set of mass-radius relationships for ocean planets with different compositions and different equilibrium temperatures, well adapted to low-density sub-Neptune planets. While it does not preclude the existence of other hydrogen-helium-rich sub-Neptunes, our model suggests that super-Earths and water-rich sub-Neptunes could belong to the same family of planets. The differences between their interiors could simply result from the variation of the water content in those planets. 

\section{Model}
\label{Model}
\subsection{Interior model}
\label{Int}

We use the internal model of solid planet developed by \cite{Br17}. It takes as inputs the planetary mass and chemical composition (Mg/Si, Fe/Si mole ratios and water mass fraction), and computes the resulting radius and internal structure of the planet \citep{Br17}. The internal structure computes the pressure $P(r)$, the temperature $T(r)$, the gravity acceleration $g(r)$, and the density $\rho(r)$ as a function of radius. These quantities are integrated following an iterative scheme until convergence is reached. Along the radius $r$ of the planet, the pressure $P(r)$ is calculated via different EOS, which are chosen depending on the material that composes the considered layer. The different layers include the core, the lower and upper mantles, the high pressure ice, and a liquid hydrosphere \citep{Br17}. To account for the effects of irradiation, as expected for the close-in population, a water phase in supercritical state has been added to the hydrosphere. For given density and temperature, the supercritical layer pressure is calculated via an EOS obtained from data generated by molecular level computer simulations that consider simple point-charge potential models to which average polarization corrections have been added \citep{Du06}. This EOS is written as:

\begin{equation}
\label{eq1}
\begin{split}
Z = \frac{PV}{RT} = 1 + \frac{BV_c}{V} + \frac{CV_c^2}{V^2} + \frac{DV_c^4}{V^4} + \frac{EV_c^5}{V^5} +\\
 \frac{FV_c^2}{V^2} \times \left(\beta + \frac{\gamma V_c^2}{V^2}\right)~\rm{exp} \left(-\frac{\gamma V_c^2}{V^2}\right),
\end{split}
\end{equation}

\noindent where $R$ = 83.14467 cm$^3$ bar/(K mol) is the universal gas constant. Parameters $B$, $C$, $D$, $E$, and $F$ in Eq. \ref{eq1} are calculated via the following equations:

\begin{equation}
\label{eq2}
B = a_1 + \frac{a_2}{T_r^2} + \frac{a_3}{T_r^3}
\end{equation}

\begin{equation}
\label{eq3}
C = a_4 + \frac{a_5}{T_r^2} + \frac{a_6}{T_r^3}
\end{equation}

\begin{equation}
\label{eq4}
D = a_7 + \frac{a_8}{T_r^2} + \frac{a_9}{T_r^3}
\end{equation}

\begin{equation}
\label{eq5}
E = a_{10} + \frac{a_{11}}{T_r^2} + \frac{a_{12}}{T_r^3}
\end{equation}

\begin{equation}
\label{eq6}
F = \frac{\alpha}{T_r^3}
\end{equation}

\begin{equation}
\label{eq7}
T_r = \frac{T}{T_c}
\end{equation}

\begin{equation}
\label{eq8}
F = \frac{RT_c}{P_c}
\end{equation}

\noindent where $T_c$ and $P_c$ are the critical temperature and critical pressure respectively. Here, $T_c$ = 647.25 K, and $P_c$ = 221.19 cm$^3$/mol. Parameters $a_1$--$a_{12}$, $\alpha$, $\beta$, and $\gamma$ valid in the 0.2--10 GPa range are summarized in Table 1. We refer the reader to the study of Duan \& Zhang (2006) for details.

\begin{table}[h]
\begin{center}  
\caption{EoS Parameters}
\smallskip         
\begin{tabular}{lc}     
\hline       
\hline       
Parameter		& Value		    					\\
\hline
$a_1$		&  4.68071541 $\times$ 10$^{-02}$ 		\\                                      	
$a_2$		& -2.81275941 $\times$ 10$^{-01}$ 		\\                                      	
$a_3$		& -2.43926365 $\times$ 10$^{-01}$ 		\\                                      	
$a_4$		&  1.10016958 $\times$ 10$^{-02}$ 		\\                                      	
$a_5$		& -3.86603525 $\times$ 10$^{-02}$ 		\\                                      	
$a_6$		&  9.30095461 $\times$ 10$^{-02}$ 		\\                                      	
$a_7$		& -1.15747171 $\times$ 10$^{-05}$ 		\\                                      	
$a_8$		&  4.19873848 $\times$ 10$^{-04}$ 		\\                                      	
$a_9$		& -5.82739501 $\times$ 10$^{-04}$ 		\\                                      	
$a_{10}$		& 1.00936000 $\times$ 10$^{-06}$ 		\\                                      	
$a_{11}$		& -1.01713593 $\times$ 10$^{-05}$ 		\\                                      	
$a_{12}$		& 1.63934213 $\times$ 10$^{-05}$ 		\\                                      	
$\alpha$		& -4.49505919 $\times$ 10$^{-02}$ 		\\                                      	
$\beta$		& -3.15028174 $\times$ 10$^{-01}$ 		\\                                      	
$\gamma$	&  1.25000000 $\times$ 10$^{-02}$ 		\\                                      	
\hline                  
\end{tabular}
\end{center}
\label{tab1}
\end{table}

 The resulting EOS (hereafter DZ06) agrees within a $\pm$0.6\% deviation with the well-known IAPWS95 formulation \citep{Wa02}, which provides an accurate EOS based on experimental data within the $\sim$0--1.0 GPa pressure range. At higher pressure, the DZ06 EOS has been shown to compute the pressure within a $\pm$1.3\% deviation up to 10.0 GPa. Above, comparisons with simulated data \citep{Du96,Du06} shows it remains within a $\pm$5.0\% deviation up to 35 GPa.

The adiabatic temperature profile within the supercritical layer depends on the Gr\"uneisen parameter, which has a strong dependence with both density and temperature. In the supercritical layer, this parameter is derived from a bilinear interpolation of a grid of data available in the python library for IAPWS standard calculation of water and steam properties\footnote{https://pypi.org/project/iapws/\#description}. This grid gives a range of Gr\"uneisen parameters for temperatures up to 10$^4$~K and supercritical water densities up to 2500~kg/m$^3$, corresponding to pressures up to $\sim$150~GPa, a value exceeding the one at the center of a 20 $\Mearth$ planet fully made of water. 
When deriving this grid, the IAPWS team focused on the behavior of the extrapolation of the analytical  formulation, and ensured it remains physically correct in domains of high pressure/temperature, which are relevant to exoplanetary interiors. The Gr\"uneisen parameter's profile is then expected to have a correct physical behavior, albeit with increasing uncertainties when going deeper in the planet. However, we find this to be of secondary importance regarding planetary radius as the Gr\"uneisen parameter is basically a proxy of thermal expansivity along pressure variations, which rapidly becomes of second order when pressure increases.

\subsection{Atmosphere model}

The atmosphere model \citep{Ma12,Ma17} takes over the hydrosphere at water column pressures lower than $\sim$1000 bar, where the H$_2$O envelope behaves more and more like a hot and dense steam atmosphere as the pressure drops. The used model is based on a $T(P)$ profile prescription \citep{Ka88} starting from the 1000-bar level (unsaturated since $T(1000\,\mathrm{bar}) > T_{\mathrm{critical}}$) upwards, assuming a dry adiabat, and switching optionally to a moist adiabat (where $T(P) = T_{\mathrm{saturation}}(P)$) if/when saturation reaches unity. Once the temperature reaches the top temperature, here set to 200 K, an isothermal radiative mesosphere $T=T_{\mathrm{top}}$ is assumed up to the 0.1~Pa topmost level. Moreover, steam is not treated as an ideal gas, and the EOS is taken instead from the NBS/NRC steam tables \citep{Ha84}. This enables a smooth transition of the $T(P)$ profile with the interior model at equilibrium temperatures of 300 and 650 K, at least. Altitudes are computed assuming hydrostatic equilibrium. Shortwave and thermal fluxes are then computed using $4$-stream approximation. Gaseous (line and continuum) absorptions are computed using the $k$-correlated method. Rayleigh opacity is also included. $T(1000\,\mathrm{bar})$ is iteratively chosen so that the thermal flux at the top of the atmosphere is equal to $\sigma {T_{\mathrm{eq}}}^4$. We finally chose the radius/altitude of the 20 mbar level as the observable, transiting radius \citep{Gr18}.

\section{Implications for the mass-radius relationships}
\label{Results}
\begin{figure}[ht!]
\includegraphics[width=\linewidth]{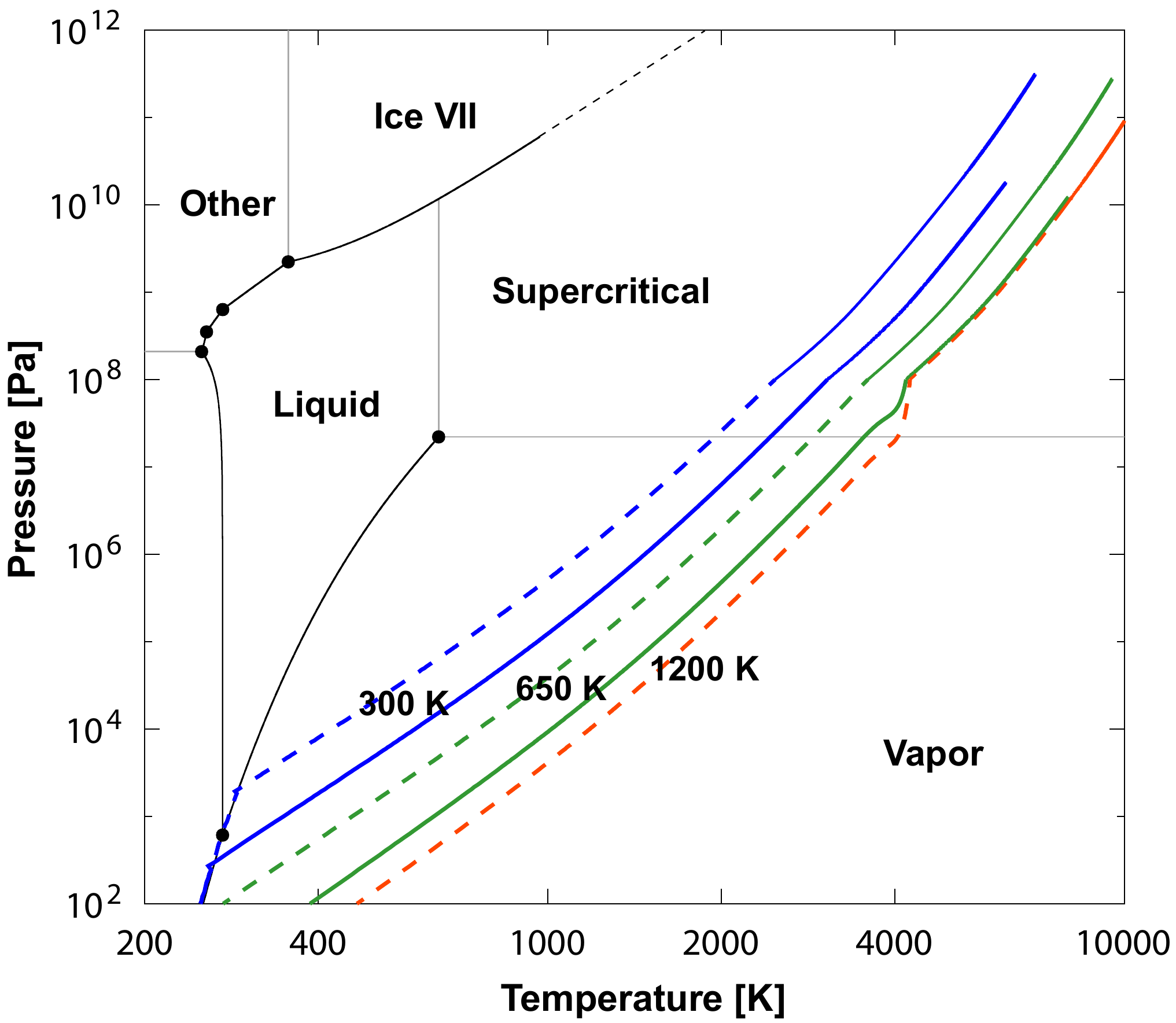}
\caption{Pressure and temperature profiles of hydrospheres of 1 and 15 $\Mearth$ supercritical ocean planets fully constituted of water with equilibrium temperatures of 300, 650, and 1200 K superimposed onto the water phase diagram. Colored solid and dashed curves correspond to 1 and 15 $\Mearth$ planets, respectively.  The black dashed line delimits a hypothetical transition between the supercritical phase and high pressure ice (see text). Dashed lines change to solid lines when the interior model takes over from the atmosphere model. The 1 $\Mearth$ case with an equilibrium temperature of 1200 K is removed because its atmosphere is hydrostatically unstable (see Fig. \ref{MR}).}
\label{adia}
\end{figure}

Figure \ref{adia} displays the pressure and temperature profiles (hereafter ($P$, $T$) profiles) of 1--15 $\Mearth$ supercritical ocean planets fully constituted of water with equilibrium temperatures of 300, 650, and 1200~K, superimposed onto the water phase diagram. With the lack of rocky cores, these planets present unphysical interiors but they have the merit to display hydrospheres over a large range of temperatures and pressures. The fluid--ice VII transition law is fitted from data between 3 and 60 GPa \citep{Fr04}. The other phase-change laws are collected from a compilation of thermodynamic data \citep{Wa02}, and from the website of the International Association for the Properties of Water and Steam (IAPWS)\footnote{http://www.iapws.org/relguide/MeltSub.html}. The ($P$, $T$) profiles expand from the base of the hydrosphere (here the center of the planet) to the top of the H$_2$O-dominated atmosphere set to 0.1~Pa. Most of the hydrospheres remain in the supercritical regime and those of smallest planets are located well below ice VII, with a fluid-ice VII transition law valid up to 60 GPa \citep{Fr04} in the water phase diagram. Beyond this pressure range, the phase change from supercritical to high pressure ices (VII or X) is neglected because the temperature/pressure region remains widely unknown in this region. 

\begin{figure}[ht!]
\includegraphics[width=\linewidth]{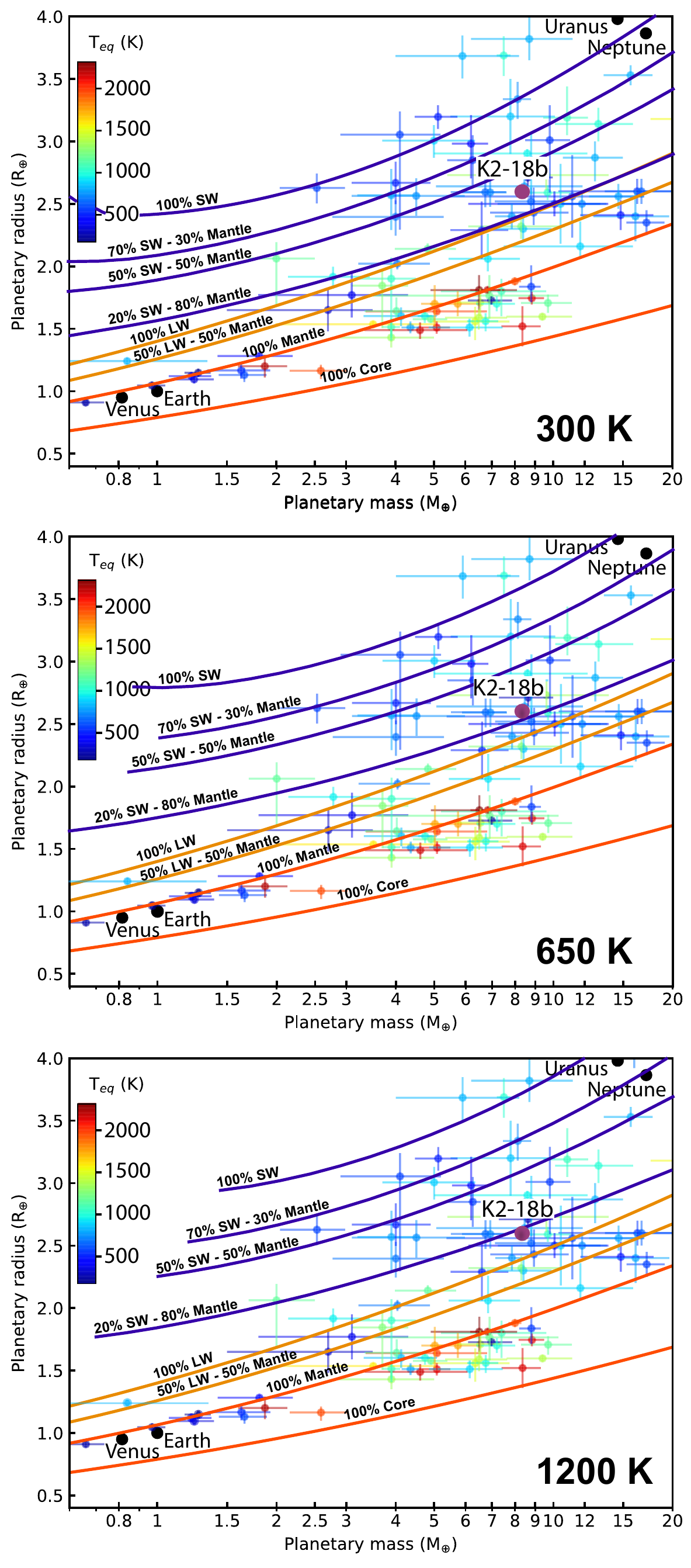}
\caption{Mass-radius diagrams determined for exoplanets with masses in the 0.6--20 $\Mearth$ range, and equilibrium temperatures of 300 K, 650 K, and 1200 K. Mass-radius curves are calculated for several planetary compositions: 100\% core and 100\% mantle (red curves), liquid water (LW) hydrosphere (brown curves) and supercritical water (SW) hydrosphere (blue curves) topping mantle-like composition interiors. Planetary data are taken from the NASA exoplanet archive and updated to 20th July 2019. Hydrostatically unstable atmospheres (defined when the altitude at 0.1 Pa tends towards infinity) around the hotter and smaller planets are excluded from the mass-radius relationships.}
\label{MR}
\end{figure}


Figure \ref{MR} represents the mass/radius relationships of supercritical ocean planets calculated in the 0.6--20 $\Mearth$ range for equilibrium temperatures of 300, 650, and 1200 K, corresponding to distances of 0.72, 0.15, and 0.04 AU from a solar-type star, respectively, assuming an Earth-like albedo. We explore different interior compositions, ranging between the two extreme and unrealistic cases of pure core and pure supercritical water (SW) compositions, to compare the effects on the resulting observables. The presence of a thick H$_2$O-dominated atmosphere generates a strong runaway greenhouse effect causing the presence of a supercritical hydrosphere, even if the equilibrium temperature of the planet is lower than the critical temperature of water ($T_{\mathrm{critical}}$ $\sim$650 K). Because the core-mantle boundary is not firmly defined at very high pressure and temperature \citep{Ha18}, we assume here the core and mantle form  a unique phase constituted of silicate rocks, including bridgmanite (Mg,Fe)SiO$_3$, and ferro-periclase, (Mg,Fe)O \citep{Br17}.

Figure 2 shows that some members of the sub-Neptunes population can be well matched by mass/radius curves corresponding to ocean planets with significant supercritical hydrospheres. For the sake of comparison, the sub-Neptune K2-38b, in which H$_2$O absorption bands have been recently observed \citep{Be19b}, is also represented in Fig. \ref{MR}. If one considers the central values of its measured mass and radius ($M_p$ = 8.63 $\Mearth$ and $R_p$ = 2.61 $\Rearth$; \cite{Be19b}), the composition of this planet cannot be explained via the use of classical super-Earths interior models, and requires an extra amount of hydrogen-helium. On the other hand, it is fully matched by models made by a rocky core-mantle toped by a supercritical water layer and a steam atmosphere irradiated at 300~K, a value close to the equilibrium temperature estimated to be $\sim$257 K \citep{Be19b}. In this case, the mass fraction of incorporated water is $\sim$37\%, a value in the same range as those derived for the water worlds in our solar system.

\section{Discussion}

Our model explores the assumption of water-rich and supercritical planets as an alternative or a complement to sub-Neptunes with hydrogen/helium gaseous envelopes. This indeed does not exclude the existence of such mini-giants but suggests that super-Earths and a fraction of sub-Neptunes could belong to the same family of planets, with differences between their interiors simply resulting from the variation in the water content. Because of the proximity to their host star, for those which are water-rich, the strong insolation associated to runaway greenhouse effect in their atmospheres generates inflated supercritical hydrospheres. As a result, they would exhibit larger radii compared to similar bodies with a very low water content located at higher distances to the star. Both \cite{Ze14} and \cite{Th16} also found that small planets with supercritical water layers could exhibit larger radii than planets of the same masses with solid or liquid hydrospheres. However, these increases were found more moderate than those derived from our model because of the absence of coupling with water-rich atmosphere models. The same effect has been recently reported in the case of irradiated super-Earths displaying water steam atmospheres in contact with rocky mantles \citep{Tu19}. However, the outcomes of this study require a new assessment via the implementation of a dedicated supercritical water EOS in the atmosphere model. Meanwhile, our model also presents some limitations. The convergence observed in Fig. \ref{adia} at the 1000-bar interface between the pressure-temperature profiles of the 15 $\Mearth$ planet at 1200 K and the 1 $\Mearth$ planet at 650 K corresponds to the use of an EOS for supercritical water in the atmosphere model different from the one employed in the interior model, and which slightly departs from its validity range.

We underline that high pressure ices such as Ice X and Ice XI \citep{Ze14} are not taken into account in our model. However, the pressure reached at the base of the hydrosphere of 20 $\Mearth$ planets including 50\% water is about 600 GPa with temperatures always greater than 4000 K. This temperature-pressure range, close to the one reached at the center of the 15 $\Mearth$ pure water planet (500 GPa) shown in Fig. \ref{adia}, is out of the stability domain of Ice X and Ice XI (see Fig. 1 of \cite{Ze14}). We therefore estimate that this effect should be marginal up to 20 $\Mearth$. Also, high temperatures achieved at the base of the hydrosphere may be sufficient to induce melting of the silicate layer. As a result, much of the volatile content of the planet may actually be dissolved within the liquid silicate mantle \citep{Ki19,Ki20}.

Interestingly, planets possessing exactly the mass and radius of Neptune could be matched by ocean planets if these latter contain $\sim$70\% of supercritical water, depending on their orbital distance and the type of their host star. Our model implies that both super-Earths and water-rich sub-Neptunes have grown from building blocks with a wide range of ice-to-rock ratios. This hypothesis is supported by recent dynamical simulations showing that the observed distributions of close-in small planets is matched by the presence of both ice-rich and dry planets \citep{Iz17,Iz19}. Water-rich small planets should have grown from ice-rich building embryos formed beyond the snowline in protoplanetary disks while dry ones would have grown from rocky embryos formed within the snowline. Ice-rich embryos would have subsequently migrated inward the disk, catalyzing the growth of purely rocky planets interior to the ice-rich ones \citep{Ra18a,Ra18b}.
 
Observations have shown that the population of small planets follows a bimodal distribution peaking at $\sim$1.3~$\Rearth$ (super-Earths) and 2.4~$\Rearth$ (sub-Neptunes), with few planets in between \citep{Fu17,Fu18}. A study, combining planetary interior models based on a mixture of rocks and ices associated with Monte Carlo simulations, has recently addressed this question \citep{Ze19}. The authors found that the 2--4 $\Rearth$ planets should contain significant amounts of H$_2$O-dominated ices in addition to rock, with perhaps more than half, by mass. They also derived that planets with radii $>$ 3$\Rearth$ would generally require the presence of a gaseous envelope. The physical assumptions formulated in this study are however questionable since highly irradiated water-rich close-in planets should display supercritical and inflated hydrospheres instead of being simply ice-rich. It has also been proposed that the gap between the super-Earths and sub-Neptunes could be accounted by the evaporation of gaseous envelopes, due to the host star irradiation \citep{Ow18,Ji18}. The same process could also happen to water-rich sub-Neptunes displaying supercritical envelopes: studies of water loss from terrestrial planets orbiting ultracool dwarfs assume that photolysis, i.e. the conversion of H$_2$O steam into H$_2$ gas, is not a limiting process in the escape mechanism \citep{Bo17}. The same assumption would remain valid in the case of stars displaying stronger FUV fluxes.

In summary, models of water-rich planets with supercritical envelopes can explain the observed physical properties of some of the planets among the small planet population. They offer an interesting alternative to current dichotomy between rocky and gaseous-rich planets. This quantitative exploration of the role of supercritical water in planetary envelopes, which underlines the importance of composition in case of strong irradiation, will be extended by the development of a model describing the planets' interior and atmosphere in a more consistent way. This work also highlights the need for improved EOS in the high pressure and high temperature regime of the water phase diagram.

\acknowledgements
O.M. and M.D. acknowledge support from CNES. We acknowledge an anonymous Referee whose comments helped improve and clarify this manuscript.


\end{document}